\documentclass[aps,showpacs,preprintnumbers,amsmath,amssymb]{revtex4}
\usepackage{dcolumn}
\usepackage[dvips]{epsfig}
\usepackage{float}
\usepackage{graphics,epsfig}
\usepackage{graphicx}
\usepackage{dcolumn}
\usepackage{bm}
\usepackage{gensymb}
\usepackage{subfigure}
\usepackage{diagbox}
\usepackage{enumerate}
\usepackage{xcolor}
\usepackage{amsmath}
\usepackage{color}
\usepackage{ulem}
\usepackage{cancel}

\begin{document}
\baselineskip=0.5 cm
\title{\bf Dynamic shadow of a black hole with a self-interacting massive complex scalar hair}

\author{ Mingzhi Wang$^{1}$\footnote{wmz9085@126.com}, Cheng-Yong Zhang$^{2}$\footnote{zhangcy@email.jnu.edu.cn}, Songbai Chen$^{3,4}$\footnote{csb3752@hunnu.edu.cn}, Jiliang Jing$^{3,4}$\footnote{jljing@hunnu.edu.cn}}
\affiliation{$ ^1$School of Mathematics and Physics, Qingdao University of Science and Technology, Qingdao, Shandong 266061, People's Republic of China \\
$ ^2$College of Physics and Optoelectronic Engineering, Jinan University, Guangzhou 510632, People's Republic of China \\
$ ^3$Institute of Physics and Department of Physics, Key Laboratory of Low Dimensional Quantum Structures
and Quantum Control of Ministry of Education, Synergetic Innovation Center for Quantum Effects and Applications,
Hunan Normal University, Changsha, Hunan 410081, People's Republic of China\\
$ ^4$Center for Gravitation and Cosmology, College of Physical Science and Technology,
Yangzhou University, Yangzhou 225009, People's Republic of China}

\begin{abstract}
\baselineskip=0.4 cm

{\bf Abstract}
We investigate the dynamic shadows of a black hole with a self-interacting massive complex scalar hair. The complex scalar field $\psi$ evolves with time $t$, and its magnitude on the apparent horizon $|\psi_{h}|$ starts from zero, undergoes a sharp rise followed by rapid oscillations, and eventually converges to a constant value. The variation in the photon sphere radius $r_{ps}$ is similar to that of the magnitude $|\psi_{h}|$. Owing to the emergence of the complex scalar hair $\psi$, the apparent horizon radius $r_{h}$ starts increasing sharply and then smoothly approaches a stable value eventually. The shadow radius $R_{sh}$ of the black hole with an accretion disk increases with time $t_{o}$ at the observer's position. In the absence of an accretion disk, the shadow radius $R_{sh}$ is larger and also increases as $t_{o}$ increases. Furthermore, we slice the dynamical spacetime into spacelike hypersurfaces for all time points $t$. For the case with an accretion disk, the variation in $R_{sh}$ is similar to that in the apparent horizon $r_{h}$, because the inner edge of the accretion disk extends to the apparent horizon. In the absence of an accretion disk, the variation in $R_{sh}$ is similar to that in the photon sphere radius $r_{ps}$, because the black hole shadow boundary is determined by the photon sphere. As the variation in $r_{ps}$ is induced by $\psi$, it can be stated that the variation in the size of the shadow is similarly caused by the change in $\psi$. Regardless of the presence or absence of the accretion disk, the emergence of the complex scalar hair $\psi$ causes the radius $R_{sh}$ of the shadow to start changing. Moreover, we investigate the time delay $\Delta t$ of lights propagating from light sources to the observer. These findings not only enrich the theoretical models of dynamic black hole shadows but also provide a foundation for testing black hole spacetime dynamics.

\end{abstract}

 \maketitle
\newpage
\section{Introduction}
The images of supermassive black holes at the center of the giant elliptical galaxy M87 and the Milky Way galaxy have been published by the Event Horizon Telescope (EHT) Collaboration \cite{fbhs1,sga1}. These images not only confirm the existence of black holes in our universe but also marks a milestone in the fields of astrophysics and black hole physics. An increasing number of scholars have begun to focus on the study of black hole images. A black hole shadow \cite{synge,sha2,lumi,sha3} is the dark silhouette in a black hole image. It appears because photons close to the black hole can be absorbed or bent, leaving a dark shadow in the observer's sky. Analyzing the characteristics of black hole images can reveal the spacetime structure of black holes, the dynamics of accretion disks, and the physical laws in strong gravitational fields. Recent studies on black hole images have largely focused on constraining black hole parameters \cite{sha9,sha8,dressed,Intcur,bhparam,obsdep,constr}, investigating dark matter \cite{polar7, drk, polar8,shadefl,shasgra}, and verifying various theories of gravity \cite{safeg,lf,sha10,fR, 2101, 2107, 2111, Jing, Jing1, Jing2023, Jing2021, Fang, Chen, Pan}. Several other aspects of black hole shadows have also been studied \cite{dkerr, fpos2, sb10,sw, swo, astro, chaotic, my, sMN, swo7, mbw, mgw, scc, shan3add, pe, halo, review, lf2, BI, zzl1, qx2, lxy1, zzl2, bir, qx4}.

Most studies on black hole images only consider static or stationary black holes. The collapse of a compact star into a black hole, perturbation of a black hole, or mergers of black holes all result in dynamic spacetime. Studies on black hole images in dynamical spacetime are extremely scarce; nevertheless, these dynamic images hold considerable significance for the study of black holes and gravitational theories. Currently, the EHT team is working on creating a movie for Sgr A* because of its rapidly changing appearance, hoping to understand the structure of black hole. We have investigated the dynamic shadows of a Schwarzschild black hole perturbed by a specific polar gravitational wave\cite{mgw} and demonstrated that the black hole shadow exhibits periodic variations, oscillating in response to the perturbation of gravitational wave. Y. P. Zhang et al have investigated the dynamic emergence of black hole shadows in the spacetime of a collapsing boson star\cite{zyp}, revealing the evolutionary behaviors of Einstein rings and the distinct formation processes of shadows.

In this study, we investigate the shadows of a black hole with a self-interacting massive complex scalar hair. The nonlinear self-interaction can induce black hole bomb phenomena beyond the famous superradiant instability\cite{bomb}. It makes the spacetime dynamic, causing the black hole shadow to change with time. The dynamic black hole shadows can serve as a powerful tool for studying the dynamical evolution of black hole spacetime. With the next generation Event Horizon Telescope(ngEHT), future astronomical observations could potentially capture dynamic black hole images, offering unprecedented insights into their temporal evolution.

The remainder of this paper is organized as follows. In Section II, we briefly introduce the dynamical spacetime of a black hole with a self-interacting massive complex scalar hair. Then, we calculate the photon sphere radius $r_{ps}$, and study how $r_{ps}$ evolves with time $t$ under the effect of the complex scalar field $\psi$. In Section III, we present the specific intensity of the direct, lensing ring and photon ring emissions of a face-on thin disk in the dynamical spacetime. With or without an accretion disk, the variations in the black hole shadow radius $R_{sh}$ with the observer's time $t_{o}$ and with the coordinate time $t$ are investigated. Moreover, we study the time delay $\Delta t=t_{o}-t_{e}$ of lights propagating from light sources to the observer in the dynamical spacetime. Finally, we present a conclusion in Section IV. In this paper, we employ the geometric units $G=c=M=1$.

\section{Dynamical spacetime of a black hole with a self-interacting massive complex scalar hair}
The black hole spacetime we consider is the Einstein-Maxwell gravity minimally coupled with a self-interacting massive complex scalar $\psi$\cite{bomb}. The Lagrangian density is
\begin{eqnarray}
\label{lds}
\mathcal{L}=R-F_{\mu\nu}F^{\mu\nu}-D^{\mu}\psi(D_{}\psi)^{\ast}-V(\psi).
\end{eqnarray}
$F_{\mu\nu}=\partial_{\mu}A_{\nu}-\partial_{\nu}A_{\mu}$ is the Maxwell field strength, where $A_{\mu}$ is the gauge potential. The gauge covariant derivative is defined as $D_{\mu}=\nabla_{\mu}-iq A_{\mu}$, where $q$ denotes the gauge coupling constant of the complex scalar field $\psi$. The potential is given by $V(\psi)=\mu^{2}|\psi|^{2}-\lambda|\psi|^{4}+\nu|\psi|^{6}$, where $\mu$ is the scalar field mass, and $\lambda, \nu$ are positive self-interaction parameters\cite{bomb}. The Einstein equations are given as
\begin{equation}
R_{\mu\nu}-\frac{1}{2}g_{\mu\nu}R=2T^{A}_{\mu\nu}+T^{\psi}_{\mu\nu},
\end{equation}
where the energy-momentum tensors are
\begin{align}
&T^{A}_{\mu\nu}=F_{\mu\rho}F_{\nu}^{\rho}-\frac{1}{4}g_{\mu\nu}F_{\rho\sigma}F^{\rho\sigma},\\
&T^{\psi}_{\mu\nu}=\frac{1}{2}(D_{\mu}\psi)^{\ast}(D_{\nu}\psi)+\frac{1}{2}(D_{\mu}\psi)(D_{\nu}\psi)^{\ast}+\frac{1}{2}g_{\mu\nu}(D^{\mu}\psi(D_{\mu}\psi)^{\ast}+V).
\end{align}
The Maxwell equations are written as
\begin{equation}
\nabla_{\mu}F^{\mu\nu}=\frac{1}{4}iq[\psi^{\ast}\nabla_{\mu}\psi-\psi(\nabla_{\mu}\psi)^{\ast}]g^{\mu\nu}.
\end{equation}
The scalar equation is written as
\begin{equation}
D^{\mu}D_{\mu}\psi=\frac{\partial V}{\partial|\psi|^{2}}\psi.
\end{equation}

The dynamic black hole with the self-interacting massive complex scalar hair in the Painlev$\acute{e}$-Gullstrand (PG) coordinates is expressed as follows:
\begin{eqnarray}
\label{dg}
ds^{2}=-[1-\zeta(t,r)^{2}]\alpha(t,r)^{2}dt^{2}+2\alpha(t,r)\zeta(t,r) dtdr+dr^{2}+r^{2}(d\theta^{2}+\sin^{2}\theta d\phi^{2}).
\end{eqnarray}
Here, $\alpha, \zeta$ are metric functions dependent on $t$ and $r$. This coordinate system remains regular at the apparent horizon $r_{h}$ where $\zeta(t, r_{h})=1$.
The gauge potential $A_{\mu}dx^{\mu}=Adt$. Introducing auxiliary variables\cite{bomb}
\begin{align}
&\Phi=\partial_{r}\psi,\\
&\Pi=\frac{1}{\alpha}(\partial_{t}\psi-iqA\psi)-\zeta\Phi,\\
&B=\frac{1}{\alpha}\partial_{r}A.
\end{align}
The Einstein equations can be reduced to
\begin{align}
&0=\partial_{r}\alpha+\frac{\alpha r \mathrm{Re}(\Pi\Phi^{\ast})}{2\zeta},\\
&0=\partial_{r}\zeta+\frac{\zeta}{2r}-\frac{r}{4\zeta}(\Pi\Pi^{\ast}+\Phi\Phi^{\ast}+2B^{2}+V)-\frac{r}{2}\mathrm{Re}(\Pi\Phi^{\ast}),\\
&0=\partial_{t}\zeta-\frac{1}{2}\alpha r\bigg[\Pi\Pi^{\ast}+\Phi\Phi^{\ast}+\bigg(\zeta+\frac{1}{\zeta}\bigg)\mathrm{Re}(\Pi\Phi^{\ast})\bigg].
\end{align}
The Maxwell equations are expresses as
\begin{align}
&0=\partial_{r}B+\frac{2B}{r}-\frac{q}{2}\mathrm{Im}(\Pi\Phi^{\ast}),\\
&0=\partial_{t}B-\frac{q}{2}\alpha\mathrm{Im}[(\zeta\Pi+\Phi)\psi^{\ast}].
\end{align}
The scalar equation becomes
\begin{align}
0=\partial_{t}\Pi-\partial_{r}[\alpha(\Pi\zeta+\Phi)]-\frac{2\alpha(\Pi\zeta+\Phi)}{r}-iA\Pi q+\alpha\psi\frac{\partial V}{\partial|\psi|^{2}}.
\end{align}
At asymptotic spatial infinity, we have
\begin{align}
\zeta=\sqrt{\frac{2M}{r}(1+O(\frac{1}{r}))},
\end{align}
where $M$ is the total mass. By numerically solving the above equations, we can obtain the complex scalar field $\psi(t,r)$, gauge potential $A(t,r)$, and metric \{$\alpha(t,r), \zeta(t,r)$\} over all positions ($r$) and all times ($t$). In this study, we set the potential $V(\psi)=|\psi|^{2}(1-\frac{|\psi|^{2}}{0.1^{2}})^{2}, M=1.09239$, the total charge $Q=0.9$, and $q=3$.

\begin{figure}[htb]
\center{\includegraphics[width=10cm ]{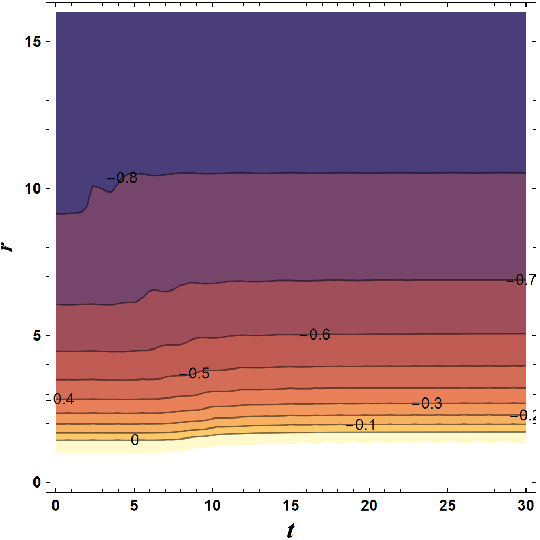}
\caption{Contour plot of $g_{tt}$ as a function of time $t$ and radial coordinate $r$. This parameter asymptotically approaches $-1$ as $r\rightarrow\infty$. The contour line of $g_{tt}=0$ corresponds to the position of the apparent horizon $r_{h}$.}
\label{gtt}}
\end{figure}

\begin{figure}[htb]
\center{\includegraphics[width=10cm ]{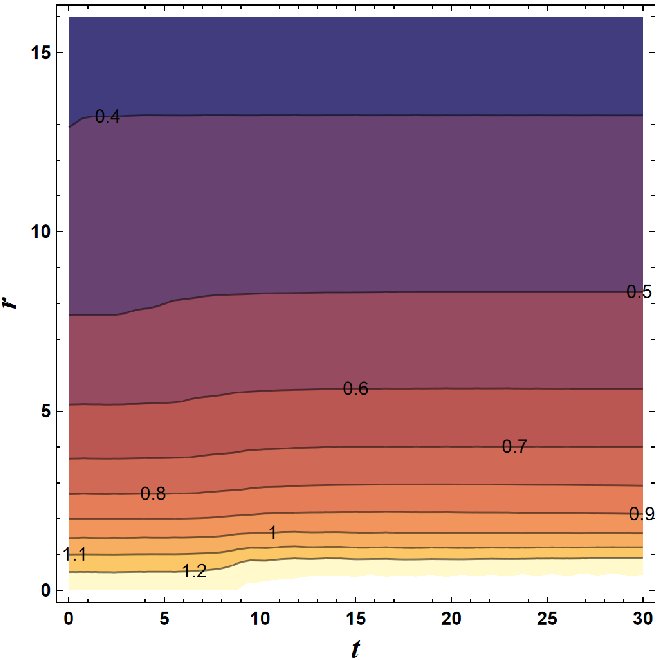}
\caption{Contour plot of $g_{tr}$ as a function of time $t$ and radial coordinate $r$. This parameter asymptotically approaches $0$ as $r\rightarrow\infty$.}
\label{ggtr}}
\end{figure}

\begin{figure}[htb]
\center{\includegraphics[width=10cm ]{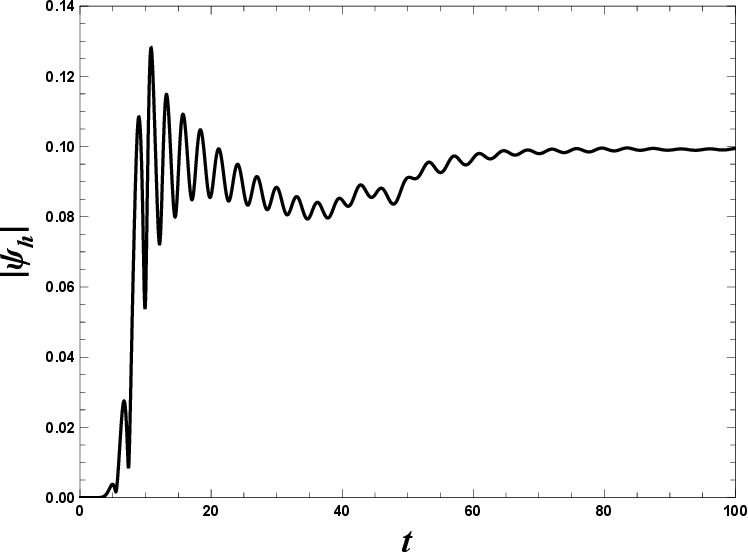}
\caption{Variation in the magnitude of the complex scalar field on the horizon $|\psi_{h}|$ with time $t$.}
\label{psi}}
\end{figure}

Figure \ref{gtt} illustrates the contour plot of the metric component $g_{tt}$ as a function of time $t$ and radial coordinate $r$. The contour line of $g_{tt}=0$ corresponds to the location of the apparent horizon $r_{h}$, which increases with time $t$. $g_{tt}$ shows little variation with $t$ and asymptotically approaches $-1$ as $r$ tends to infinity. Figure \ref{ggtr} displays a contour plot of $g_{tr}$ as a function of $t$ and $r$. $g_{tr}$ also shows little variation with $t$ and asymptotically approaches $0$ as $r$ tends to infinity. The metric of this dynamic black hole describes an asymptotically flat spacetime. The complex scalar field $\psi(t, r)$ of this black hole is a function of $t$ and $r$. Figure \ref{psi} depicts how the magnitude of the complex scalar field on the horizon $|\psi_{h}|$ changes with time. Initially, $|\psi_{h}|=0$, and at this point, the black hole has not yet acquired scalar hair. Then there is a sharp rise followed by rapid oscillations. As time progresses, the oscillations gradually dampen, and then, the magnitude tends toward stabilization.

The Lagrangian $\mathcal{L}$ of photon propagation in the  spacetime can be characterized by
\begin{equation}
\label{lagr}
\mathcal{L}=\frac{1}{2} g_{\mu\nu}\dot{x}^{\mu}\dot{x}^{\nu}=\frac{1}{2}[-(1-\zeta^{2})\alpha^{2}\dot{t}^{2}+2\alpha\zeta \dot{t}\dot{r}+\dot{r}^{2}+r^{2}(\dot{\theta}^{2}+\sin^{2}\theta \dot{\phi}^{2})]=0.
\end{equation}
As the variable $\phi$ is not present in $\mathcal{L}$, $p_{\phi}$ is a conserved quantity in the motions of photons and can be expressed as angular momentum
\begin{equation}
\label{pj}
L=p_{\phi}=\frac{\partial\mathcal{L}}{\partial\dot{\phi}}=r^{2}\sin^{2}\theta \dot{\phi}.
\end{equation}
As the spacetime is spherically symmetric, without any loss of generality, we investigate the photon motion in the plane $\theta=\pi/2$. The Lagrangian $\mathcal{L}$(\ref{lagr}) can be rewritten as
\begin{equation}
\label{lagrr}
\mathcal{L}=\frac{1}{2}[-(1-\zeta^{2})\alpha^{2}\dot{t}^{2}+2\alpha\zeta \dot{t}\dot{r}+\dot{r}^{2}+r^{2}\dot{\phi}^{2}]=0.
\end{equation}
The photon sphere is closely associated with the shadows of black holes, satisfying
\begin{equation}
\label{r12}
\dot{r}=0\;\;\;\;\;\;\; and \;\;\;\;\;\;\;\ddot{r}=0,
\end{equation}
where the dot $\cdot$ denotes the derivative with respect to the affine parameter. According to $\mathcal{L}=0$(\ref{lagrr}) and $\dot{r}=0$, one can obtain
\begin{align}
\label{r1}
-(1-\zeta^{2})\alpha^{2}\dot{t}^{2}+r^{2}\dot{\phi}^{2}=-(1-\zeta^{2})\alpha^{2}\dot{t}^{2}+\frac{L^{2}}{r^{2}}=0.
\end{align}
The equations of geodesic motion in this spacetime are
\begin{align}
&\ddot{t}=-\left(\frac{1}{2}g^{tt}g_{tt,t}+\frac{1}{2}g^{tr}\left(2g_{tr,t}-g_{tt,r}\right)\right)\dot{t}^{2}-g^{tt}g_{tt,r}\dot{t}\dot{r}-g^{tt}g_{tr,r}\dot{r}^{2}+g^{tr}r\dot{\theta}^{2}+g^{tr}\frac{L^{2}}{r^{3}\sin^{2}\theta},  \label{cdx} \\
&\ddot{r}=-\left(\frac{1}{2}g^{tr}g_{tt,t}+g_{tr,t}-\frac{1}{2}g_{tt,r}\right)\dot{t}^{2}-g^{tr}g_{tt,r}\dot{t}\dot{r}-g^{tr}g_{tr,r}\dot{r}^{2}+r\dot{\theta}^{2}+\frac{L^{2}}{r^{3}\sin^{2}\theta}, \label{cdx1}\\
&\ddot{\theta}=\frac{L^{2}\cos\theta}{r^{4}\sin^{3}\theta}-2\frac{1}{r}\dot{r}\dot{\theta}, \label{cdx2}\\
&\dot{\phi}=\frac{L}{r^{2}\sin^{2}\theta}  \label{cdx3}.
\end{align}
The radius of the photon sphere $r_{ps}$ can be determined by solving the combined conditions derived from Eqs. (\ref{r12}), (\ref{r1}), and (\ref{cdx1}), which require that $r_{ps}$ satisfies
\begin{align}
\label{r22}
g^{tr}g_{tt,t}+2g_{tr,t}-g_{tt,r}+\frac{1}{r}g_{tt}=0.
\end{align}

\begin{figure}[htb]
\center{\includegraphics[width=10cm ]{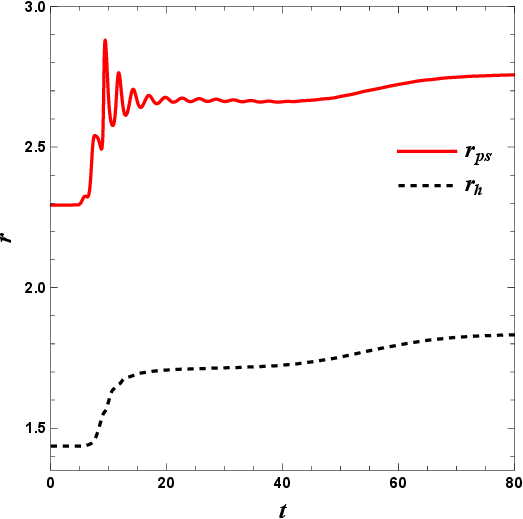}
\caption{Variations in the apparent horizon $r_{h}$ (marked by a black dashed line) and photon sphere $r_{ps}$ (marked by a red solid line) with respect to the variable $t$.}
\label{rhr}}
\end{figure}

Figure \ref{rhr} illustrates the variations in photon sphere radius $r_{ps}$ (marked by a red solid line) and the apparent horizon radius $r_{h}$ (marked by a black dashed line) with respect to time $t$. The radius of the photon sphere $r_{ps}$ remains constant initially, then experiences a significant and rapid increase subsequently, followed by some fluctuations, and finally stabilizes. The variation in the photon sphere radius $r_{ps}$ exhibits a striking similarity to that of the magnitude $|\psi_{h}|$ as illustrated in Fig.\ref{psi}. The complex scalar field $\psi$ is capable of inducing a variation in the photon sphere radius $r_{ps}$. Moreover, it will give rise to a dynamically evolving black hole shadow and endow it with new features. The radius of the apparent horizon $r_{h}$ also remains constant initially, then increases rapidly, and smoothly approaches a stable value eventually. The appearance of the complex scalar hair $\psi$ causes an increase in the apparent horizon $r_{h}$.

\section{Dynamic shadows of a black hole with a self-interacting massive complex scalar hair}
\begin{figure}[htb]
\center{ \includegraphics[width=8cm ]{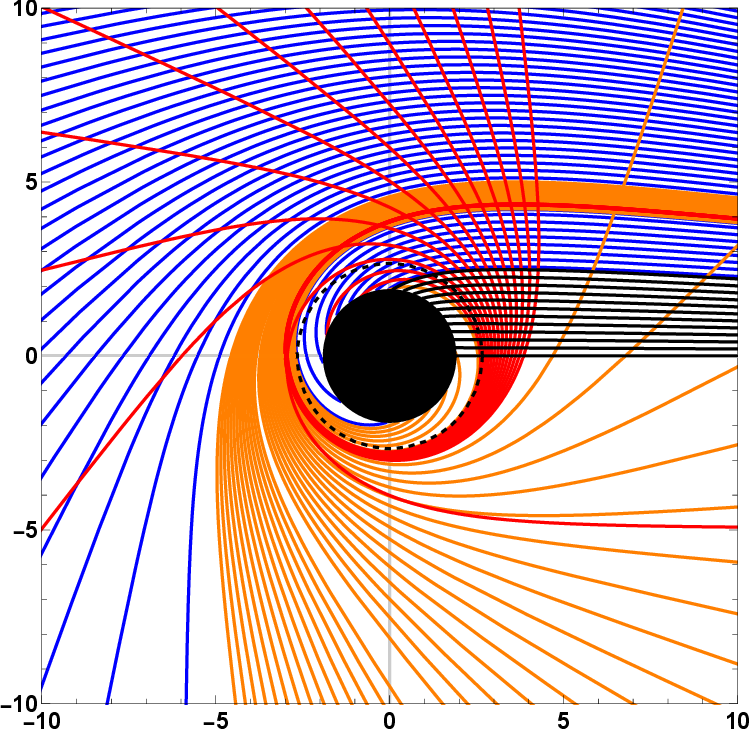}
\caption{Light rays corresponding to shadow (black), direct emission (blue), lensing rings (orange), and photon rings (red). The black disk represents the black hole, and the observer is positioned to its right at a distance of $r_{o}=50$. The accretion disk is viewed face-on by the observer.}
\label{gj}}
\end{figure}

\begin{figure}[htb]
\center{  \includegraphics[width=14.2cm ]{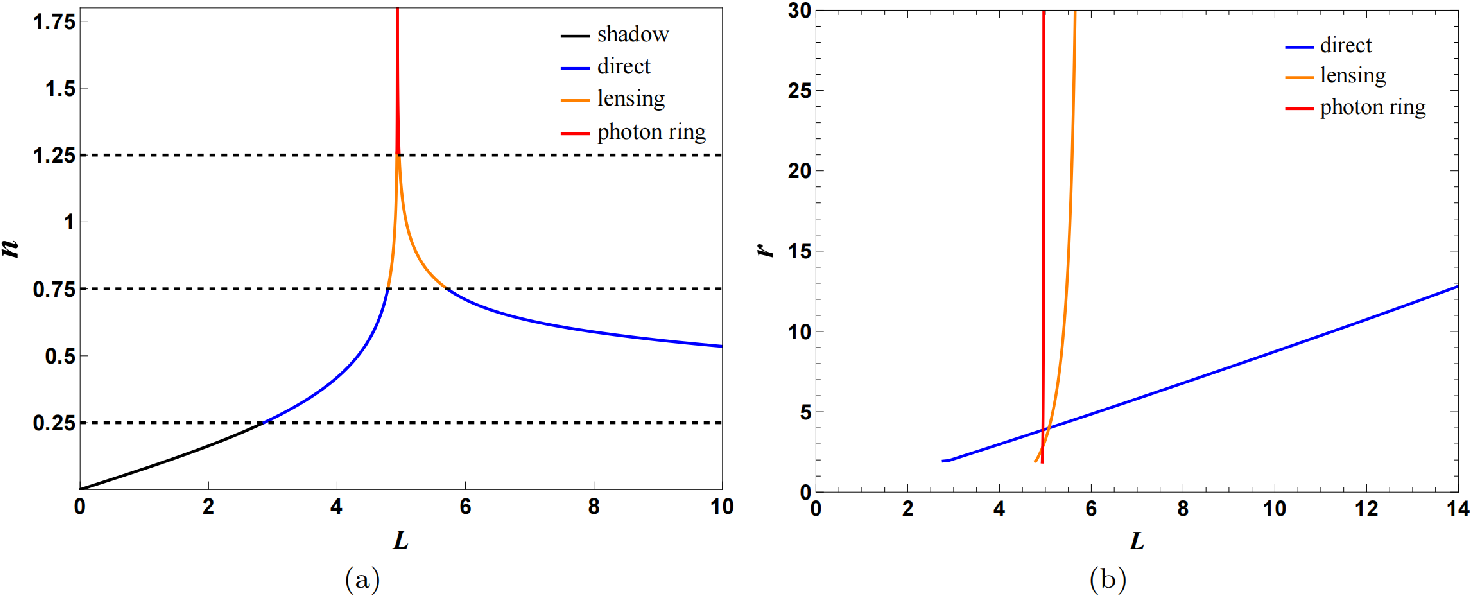}
\caption{(a)Variation in the total number $n$ with the angular momentum $L$. The black hole shadow satisfies the condition $n<1/4$ (black); the direct emission satisfies $1/4\leq n<3/4$ (blue); the lensing ring satisfies $3/4\leq n<5/4$ (orange); the photon ring satisfies $n\geq5/4$ (red). (b)Transfer functions $r_{m}(L)$ for the direct emission, the lensing ring, and the photon ring. The blue, orange, and red rays represent the radial coordinates of the first ($m=1$), second ($m=2$), and third ($m=3$) intersections with the thin disk, respectively.}
\label{dlp}}
\end{figure}

In this section, we calculate the shadows of the black hole with a self-interacting massive complex scalar hair using the backward ray-tracing method\cite{lf,sb10,pe,halo,review,sha2,sw,swo,astro,chaotic,my,sMN,swo7,mbw,mgw,scc,dkerr}. The light rays are assumed to evolve backward in time from the observer. Accordingly, it is necessary to solve the null geodesic equations numerically (\ref{cdx}-\ref{cdx3}). Figure \ref{gj} depicts the light rays from the observer to the vicinity of the black hole. The black disk represents the black hole, and the observer is positioned to its right at a distance of $r_{o}=50$. The initial time for the reverse evolution of light rays at the position of the observer is $t_{o}=170$. We consider an optically and geometrically thin accretion disk, facing directly toward the observer. Owing to the dynamical spacetime, we only consider the accretion disk extending to the horizon. In this figure, the black light rays directly evolve into the black hole, according to the black hole shadow. The size of the black hole shadow depends on the inner edge of the accretion disk. The blue rays cross the equatorial plane at most once, corresponding to direct emission; the orange rays cross the equatorial plane twice, corresponding to lensing rings; the red rays cross the equatorial plane at least three times, corresponding to photon rings\cite{bpl12,bpl}. The total number of orbits is defined as $n=\phi/2\pi$\cite{bpl12,bpl}. Figure \ref{dlp}(a) shows the variation in the total number $n$ with the angular momentum $L$. Here, the black hole shadow satisfies the condition $n<1/4$ (black); the direct emission satisfies $1/4\leq n<3/4$ (blue); the lensing ring satisfies $3/4\leq n<5/4$ (orange); the photon ring satisfies $n\geq5/4$ (red). Figure \ref{dlp}(b) shows the transfer functions $r_{m}(L)$ for the direct emission, lensing ring, and photon ring. The blue, orange and red rays represent the radial coordinates of the first ($m=1$), second ($m=2$), and third ($m=3$) intersections with the thin disk, respectively.

\begin{figure}[htb]
	\center{\includegraphics[width=10cm ]{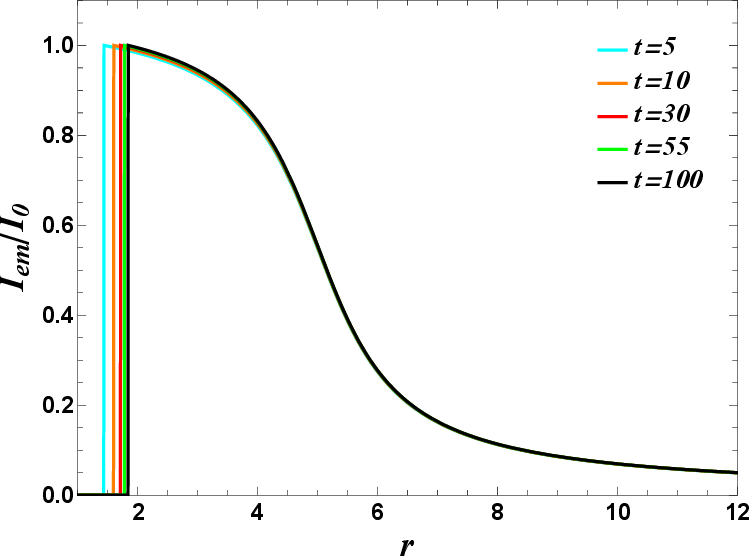}
		\caption{Variation in the emitted specific intensity $I_{\mathrm{em}}(t,r)$ at different times $t$ as a function of radial coordinate $r$.}
		\label{lem}}
\end{figure}

\begin{figure}[htb]
\center{ \includegraphics[width=15cm
]{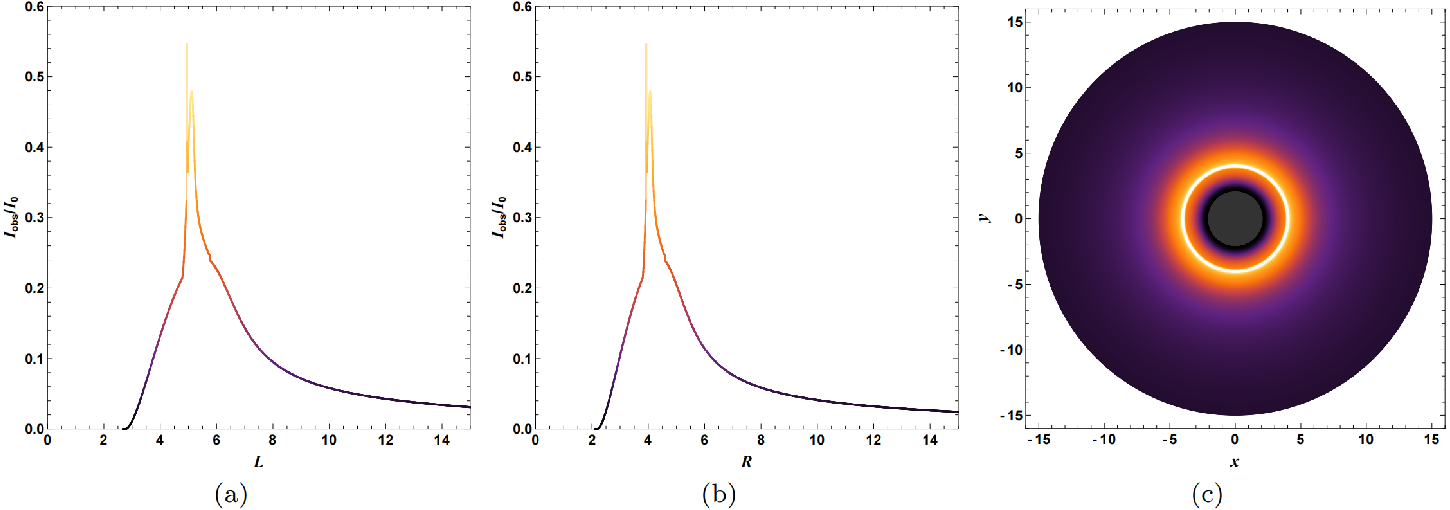}
\caption{Observed appearance of a geometrically and optically thin accretion disk, viewed from a face-on by the observer with radius $r_{o}=50$ and time $t_{o}=170$. (a)Profiles of the observed intensity $I_{obs}$ as a function of the angular momentum $L$. (b)Profiles of the observed intensity $I_{obs}$ as a function of the radial coordinate $R=\sqrt{x^{2}+y^{2}}$ in the observer's sky. (c)Density plot of the observed intensity $I_{obs}$ in the black hole image, in which the gray region represents the black hole shadow.}
\label{sh}}
\end{figure}

In the spacetime of the dynamic black hole with a self-interacting massive complex scalar hair, we consider that the geometrically and optically thin accretion disk extends to the horizon, and the emitted specific intensity is
\begin{equation}
\label{lemgs}
I_{\mathrm{em}}(t,r)=
\begin{cases}
I_0\frac{\frac{\pi}{2}-\tan^{-1}(r-5)}{\frac{\pi}{2}-\tan^{-1}[r_{\mathrm{h}}(t)-5]},& r>r_{\mathrm{h}}\\
0, & r \leq r_{\mathrm{h}}.
\end{cases}
\end{equation}
The emitted specific intensity $I_{\mathrm{em}}(t,r)$ changes with time owing to the changing apparent horizon radius $r_{h}(t)$, as shown in Eq.(\ref{lemgs}). As the accretion disk extends to the horizon, the inner radius of the accretion disk increases with time as the apparent horizon radius $r_{h}(t)$ increases, as shown in Fig.\ref{lem}. Moreover, the specific intensity received by the observer is
\begin{align}
\label{iobs}
I_{\text{obs}} = \sum_{m} g^4 I_{em}|_{r = r_m},
\end{align}
where $r_{m}$ is the radial coordinate of the $m^{th}$ intersection with the accretion disk plane\cite{bpl}. $g=\nu_{o}/\nu_{e}$ is the redshift factor, $\nu_{e}$ is the photon frequency as measured in the rest-frame of the emitter, and $\nu_{o}$ is the observed photon frequency. In this dynamical spacetime, the redshift factor $g$ can be rewritten as
\begin{align}
\label{redsh}
g = \frac{\nu_{o}}{\nu_{e}}=\frac{p_{\mu }u_{o}^{\mu}}{p_{\nu }u_{e}^{\nu}},
\end{align}
where $p_{\mu}$ is the four-momentum of the photon, $u_{o}^{\mu}=(1/\sqrt{-g_{tt}|_{(t_{o},r_{o})}}, 0, 0, 0)$ is the four-velocity of the static observer, and $u_{e}^{\mu}=(1/\sqrt{-g_{tt}|_{(t_{e},r_{e})}}, 0, 0, 0)$ is the four-velocity of the accreting gas emitting the radiation. Figure \ref{sh} shows the observed appearance of the accretion disk, viewed face-on by the observer with radius $r_{o}=50$ and time $t_{o}=170$. In other words, this is the image of the black hole captured by the observer at the local time $t_{o}=170$. Figure \ref{sh} (a) and (b) show the profiles of the observed intensity $I_{obs}$ as a function of the angular momentum $L$ and radial coordinate $R=\sqrt{x^{2}+y^{2}}$ in the observer's sky, respectively. Here, $(x, y)$ are the celestial coordinates. Figure \ref{sh} (c) shows the density plot of the observed intensity $I_{obs}$ in the observer's sky, in which the gray region represents the black hole shadow. Owing to the redshift factor $g$, the observed intensity $I_{obs}$ is not similar to the emitted specific intensity $I_{em}$, which arises from the horizon, and increases from direct emission to the lensing ring, and reaches a peak in the photon ring as $L$ increases. Then, it declines as $L$ continues to increase. The direct emission dominates a significant portion, constituting the major part in the black hole image. The lensing ring is much brighter than direct emission and forms a brilliant, glowing halo in the black hole image. Although the photon ring exhibits the highest brightness, it is so narrow that its contribution can be ignored.

\begin{figure}[htb]
\center{ \includegraphics[width=15cm ]{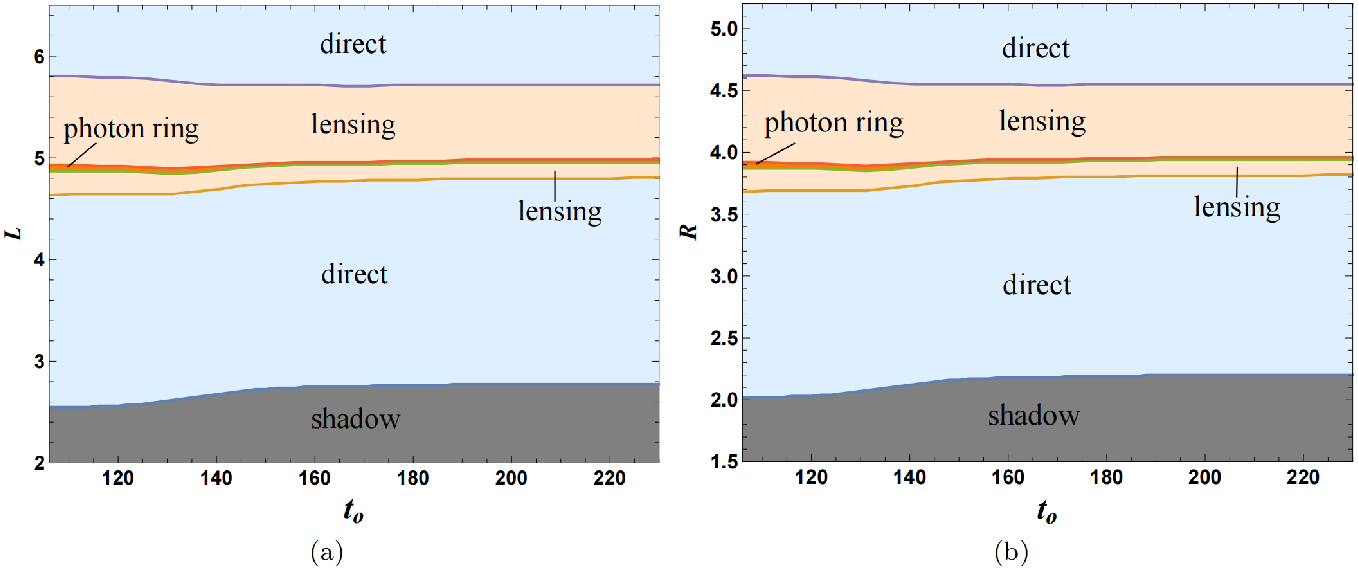}
\caption{Variations in the range of angular momentum $L$ and $R$ for the black hole shadow, direct emission, lensing ring, and photon ring with the observer time $t_{o}$. Here, the observer is located at $r_{o}=50$.}
\label{dlprl}}
\end{figure}

In the dynamical spacetime, the black hole shadow, direct emission, lensing ring, and photon ring are also dynamic with time. The variations in the range of the angular momentum $L$ and $R$ for the black hole shadow, direct emission, lensing ring, and photon ring with the observer time $t_{o}$ are shown in Fig.\ref{dlprl}. The observer is located at $r_{o}=50$. The shadow range increases with time $t_{o}$. The range of the direct emission is divided into two parts, and the entire range remains almost invariant with respect to $t_{o}$. The range of the lensing ring is also divided into two parts, but the entire range decreases with time $t_{o}$. However, the range of the photon ring is narrow and decreases with time $t_{o}$.

\begin{figure}[htb]
\center{ \includegraphics[width=15cm ]{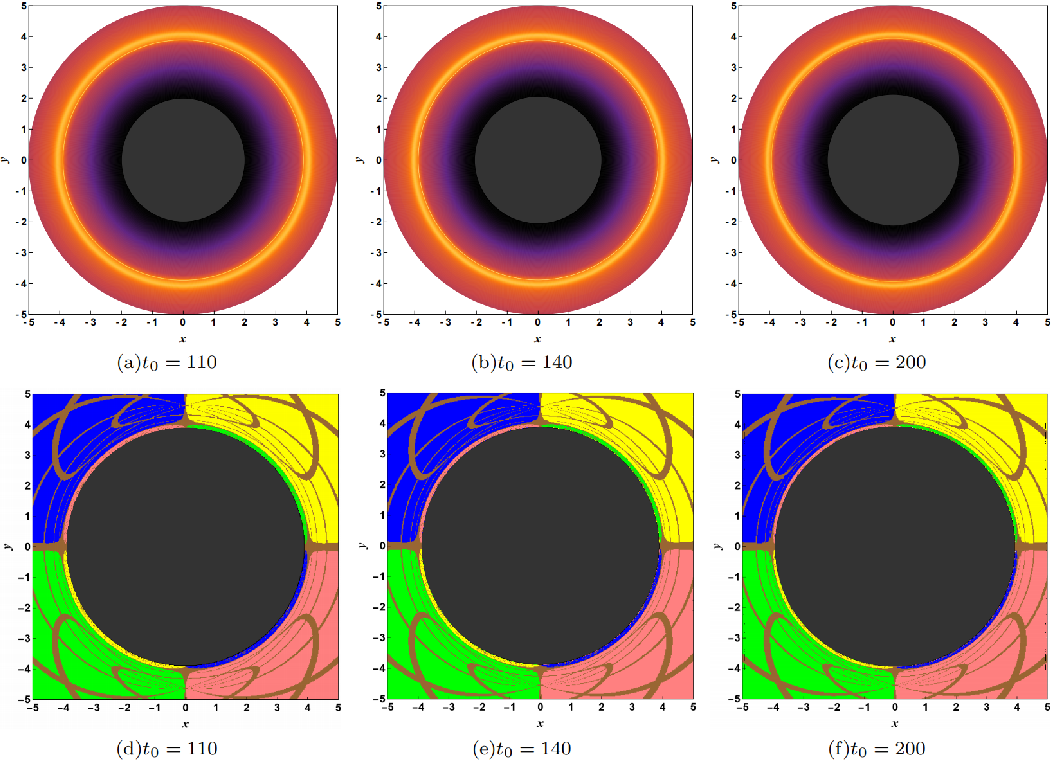}
\caption{Black hole shadows observed at $t_{o}=110, 140$, and $200$ with $r_{o}=50$. The upper row shows the images of the black hole with an accretion disk. The bottom row illustrates the black hole shadow images without an accretion disk. The gray inner regions represent the black hole shadows.}
\label{rsh50}}
\end{figure}

\begin{figure}[htb]
\center{ \includegraphics[width=15cm ]{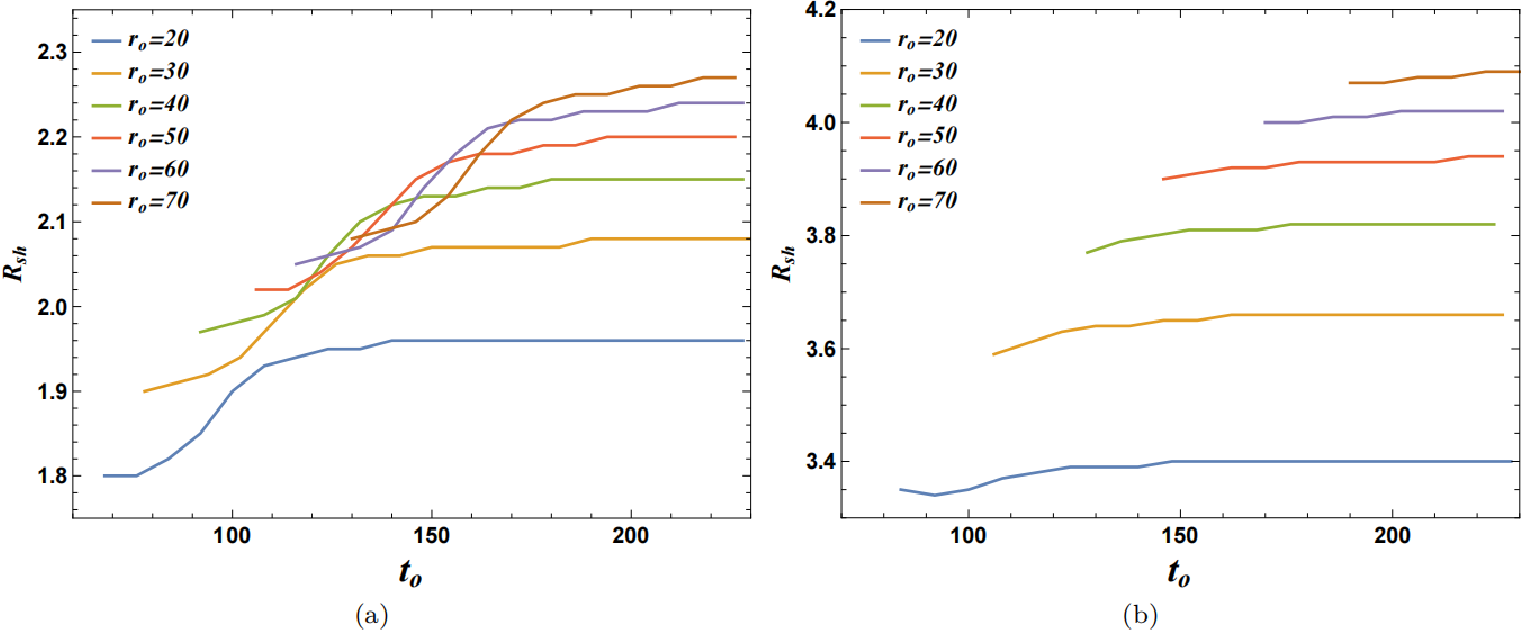}
\caption{Variations in the black hole shadow radius $R_{sh}$ with observer time $t_{o}$ for different observer distances $r_{o}$. (a)Shadow radius $R_{sh}$ for this dynamic black hole with an accretion disk. (b)Shadow radius $R_{sh}$ without an accretion disk.}
\label{rsh}}
\end{figure}

Figure \ref{rsh50} shows the black hole images observed at $t_{o}=110, 140$, and $200$ with $r_{o}=50$. The upper row of this figure shows the images of the black hole with an accretion disk. The gray inner region is the black hole shadow, whose radius increases as $t_{o}$ increases, although the increase is small. The shadow radius is approximately $2.15$. The bottom row of this figure illustrates the black hole shadows without an accretion disk. The gray region increases in size with the increase in $t_{o}$. The colored region represents the image of the spherical background light source, consistent with the configuration described in Ref.\cite{sMN}. The radius of the black hole shadow without an accretion disk is larger than that with an accretion disk, measuring approximately $3.95$. Figure \ref{rsh} illustrates the variations in the black hole shadow radius $R_{sh}$ with observer time $t_{o}$ for different observer distances $r_{o}$. Figure \ref{rsh} (a) and (b) correspond to the black hole images with and without an accretion disk, respectively. In either case, the shadow radius $R_{sh}$ of this dynamic black hole increases with observer time $t_{o}$. In the case with an accretion disk, the shadow radius $R_{sh}$ increases as the observer distance $r_{o}$ increases only when $t_{o}$ is sufficiently large. In the absence of an accretion disk, the shadow radius $R_{sh}$ is larger and increases as $r_{o}$ increases. We did not start the observer time $t_{o}$ from $0$ because of the time delay for light rays to propagate from the light source to the observer. The light rays observed in the black hole image are emitted at different times from light sources. If $t_{o}$ is too short, for some light rays (particularly those near the shadow), the backward-evolved time $t$ might be less than $0$. Moreover, times $t<0$ are beyond the scope of our consideration. From Fig.\ref{rhr}, one can observe that the apparent horizon $r_{h}$ and photon sphere $r_{ps}$ approach an asymptotic stable state as the time $t$ increases. Therefore, the variations in the shadow radius $R_{sh}$ with $t_{o}$ in Fig.\ref{rsh50} and Fig.\ref{rsh} are small.

\begin{figure}[htb]
\center{ \includegraphics[width=15cm ]{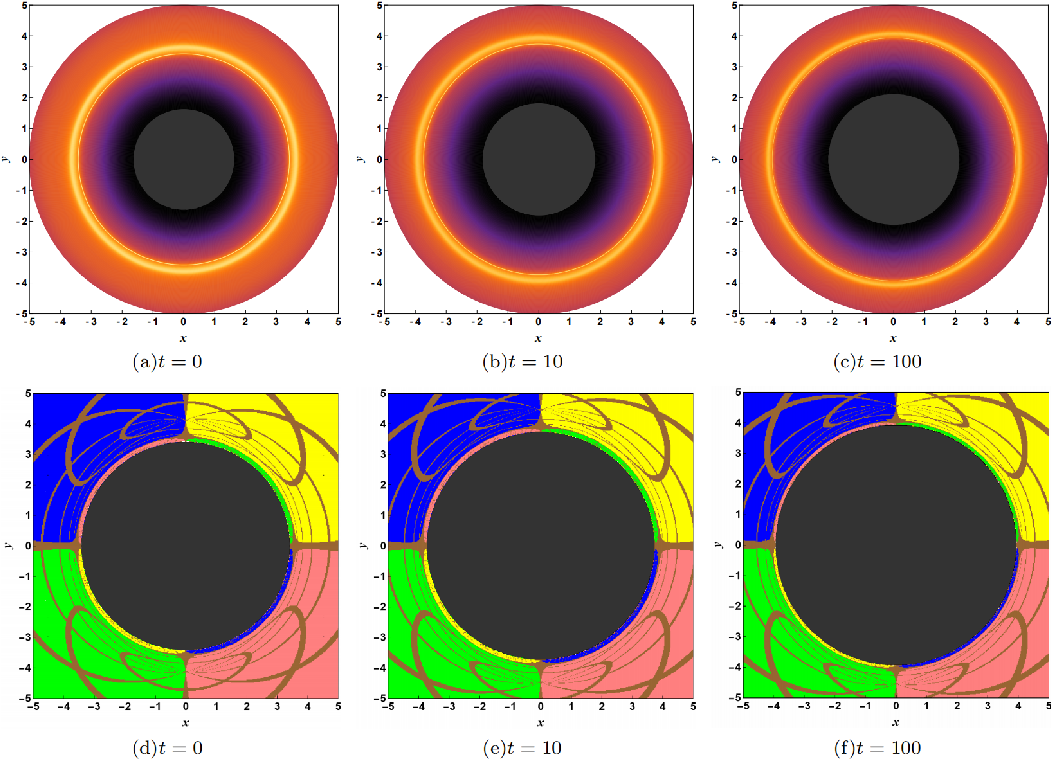}
\caption{Black hole images with and without an accretion disk when $t=0, 10$, and $100$. The upper row shows the images of the black hole with an accretion disk. The bottom row illustrates the black hole shadow images without an accretion disk.}
\label{ysh50} }
\end{figure}

\begin{figure}[htb]
\center{ \includegraphics[width=16cm ]{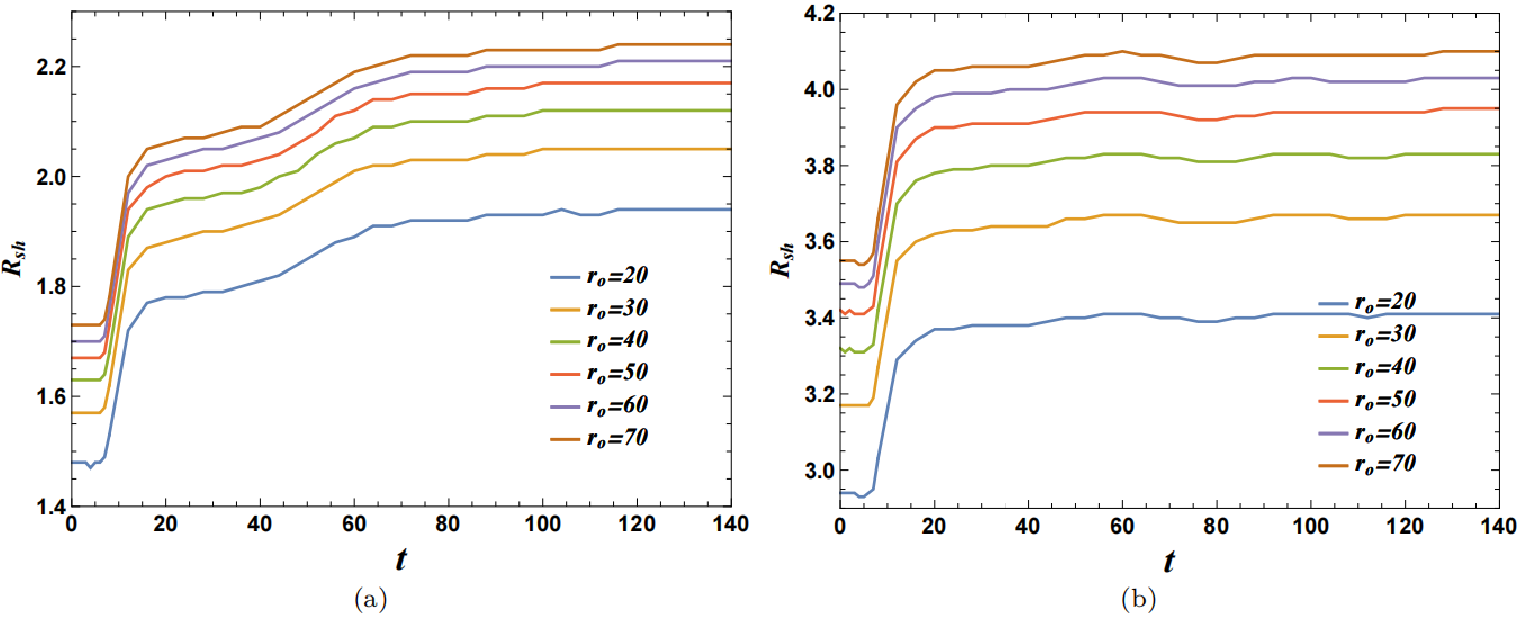}
\caption{(a)Variation in the shadow radius $R_{sh}$ for the dynamic black hole with an accretion disk. (b)Variation in the shadow radius $R_{sh}$ for the dynamic black hole without an accretion disk.}
\label{ywrsh}}
\end{figure}

To investigate the characteristics of the black hole shadow at every time point, we slice the dynamical spacetime into spacelike hypersurfaces at different times $t$. At every spacelike hypersurface, the time $t$ is fixed. We can study the black hole shadow in the spacelike hypersurface at any moment. Figure \ref{ysh50} shows the black hole shadows with $t=0, 10$, and $100$. The upper row shows the black hole images with an accretion disk, and the bottom row illustrates the black hole shadows without an accretion disk. The radius $R_{sh}$ of the black hole shadow increases as $t$ increases. Moreover, the variation in the shadow radius $R_{sh}$ is much larger than that in Fig.\ref{rsh50}. The variations in the black hole shadow radius $R_{sh}$ over time $t$ for different values of $r_{o}$ are illustrated in Fig.\ref{ywrsh}. Figure \ref{ywrsh} (a) and (b) correspond to the black hole images with and without an accretion disk, respectively. For the case with an accretion disk, the black hole shadow radius $R_{sh}$ shows no change at the beginning of time $t$ and then increases rapidly. As time $t$ progresses, it exhibits a trend of gradual stabilization. In the absence of an accretion disk, $R_{sh}$ also shows no change initially and then shows a significant increase. As time $t$ increases, it undergoes damped oscillations before finally converging to a steady state. In either case, $R_{sh}$ is larger for larger $r_{o}$. Comparing with Fig.\ref{rhr}, it can be observed that the variation in $R_{sh}$ for the case with an accretion disk is similar to that of the apparent horizon $r_{h}$, whereas the variation in $R_{sh}$ for the case without an accretion disk is similar to that of the photon sphere $r_{ps}$. This is because that the shadow boundary of the black hole with an accretion disk is determined by the inner edge of the accretion disk, which extends to the apparent horizon. For the case without an accretion disk, the shadow boundary of the black hole is determined by the photon sphere. As the variation in $r_{ps}$ is induced by the change in the complex scalar field $\psi$, it can be stated that the variation in the size of the shadow is similarly caused by the change in $\psi$. Furthermore, regardless of the presence or absence of the accretion disk, the emergence of the complex scalar hair $\psi$ causes the radius $R_{sh}$ of the shadow to start changing.

\begin{figure}[htb]
\center{ \includegraphics[width=15cm
]{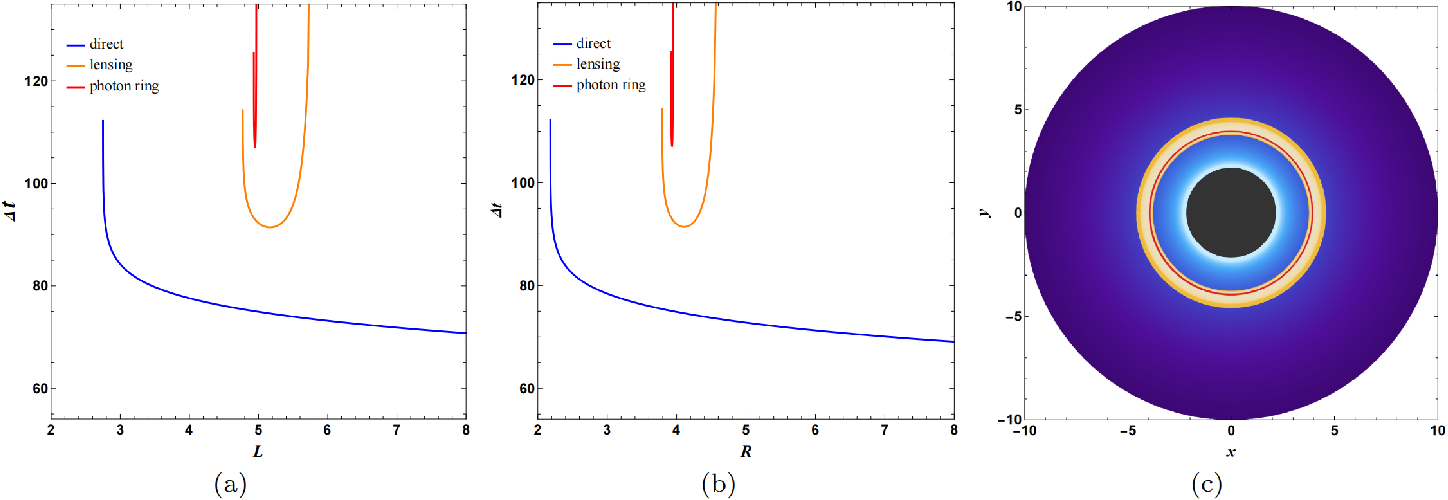}
\caption{Time delay $\Delta t=t_{o}-t_{s}$ of lights propagating from the accretion disk to the observer. (a)Variation in the time delay $\Delta t$ with the angular momentum $L$. (b)Variation in the time delay $\Delta t$ with the radius $R$ on the observer's screen. (c)Density plot of the time delay $\Delta t$ of lights in the black hole image, in which the gray inner region represents the black hole shadow.}
\label{txsh}}
\end{figure}

As the spacetime of black hole with a self-interacting massive complex scalar hair is dynamical, it is essential to investigate the time delay of light rays in the black hole image. Figure \ref{txsh} shows the time delay $ \Delta t=t_{o}-t_{e}$ of light propagating from the accretion disk to the observer. Here, the observer time $t_{o}=170$, and the radius coordinate of the observer $r_{o}=50$. Figure \ref{txsh} (a) and (b) show the variation in the time delay $\Delta t$ with the angular momentum $L$ and radius $R$ on the observer's screen, respectively. Blue, orange, and red lines represent the time delays $\Delta t$ of light from the direct emission, lensing ring, and photon ring, respectively. On average, the time delay $\Delta t$ of light rays from the photon ring is the longest, followed by that from the lensing ring, and the shortest is that from the direct emission. The time delay $\Delta t$ in the direct emission decreases with the increase in $L$ and $R$. The time delay $\Delta t$ in the lensing ring and photon ring first decreases and then increases as $L$ or $R$ increases. This is because the light rays with small $L$ originate from the accretion disk close to the horizon and are subject to stronger gravity. The light rays with large $L$ originate from the exterior of the accretion disk and require more time to traverse longer trajectories. Figure \ref{txsh} (c) shows the density plot of the time delay $\Delta t$ of light in the black hole image, in which the gray inner region represents the black hole shadow.

\begin{figure}[htb]
\center{ \includegraphics[width=15cm
]{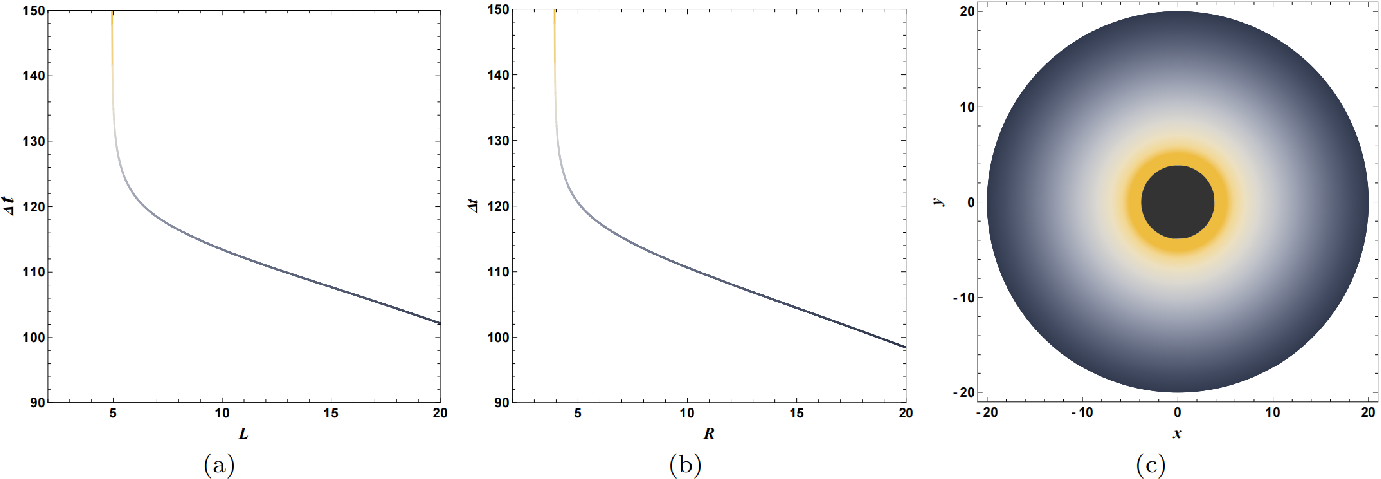}
\caption{Time delay $\Delta t=t_{o}-t_{s}$ of lights propagating from a spherical background light source to the observer. (a)Variation in the time delay $\Delta t$ with the angular momentum $L$. (b)Variation in the time delay $\Delta t$ with the radius $R$ on the observer's screen. (c)Density plot of the time delay $\Delta t$ of lights in the black hole image, in which the gray inner region represents the black hole shadow.}
\label{tsh}}
\end{figure}

In the absence of an accretion disk, light propagates from a spherical background light source with radius $r_{s}=50$ to the observer. Figure \ref{tsh} illustrates the time delay $\Delta t=t_{o}-t_{s}$ of light propagating from the light source to the observer. Here, the observer is located at $r_{o}=50$, and the observer time $t_{o}=170$. Figure \ref{tsh} (a) and (b) show the variation in the time delay $\Delta t$ with the angular momentum $L$ and radius $R$ on the observer's screen, respectively. Initially, when $L$ is small, the time delay $\Delta t$ approaches infinity because the corresponding light rays spiral asymptotically toward the photon sphere, where they can orbit the black hole multiple times. Then, the time delay $\Delta t$ decreases as $L$ or $R$ increases. Figure \ref{tsh} (c) shows the density plot of the time delay $\Delta t$ of light in the black hole image, in which the gray inner region represents the black hole shadow.
\section{Conclusion}
In this study, we investigated the dynamic shadows of a black hole with a self-interacting massive complex scalar hair. We first provided a brief review of the spacetime of the dynamic black hole. The complex scalar field $\psi$ evolves with time $t$. At the initial time $t$, the magnitude of the complex scalar field on the horizon is given by $|\psi_{h}|=0$, indicating that the black hole has not yet acquired scalar hair. Subsequently, $|\psi_{h}|$ exhibits a sharp rise, followed by rapid oscillations, and finally converges to a stable value. Influenced by the complex scalar field $\psi$, the variation in the photon sphere radius $r_{ps}$ is similar to that of the magnitude $|\psi_{h}|$. Owing to the emergence of the complex scalar hair $\psi$, the apparent horizon radius $r_{h}$ starts increasing sharply and then smoothly approaches a stable value eventually. Then, we studied the direct emission, lensing ring, and photon ring of a face-on thin disk in the dynamical spacetime. We calculated the specific intensity of light rays received by the observer from an accretion disk, and show the observed appearance of the dynamic black hole shadow. For the case with an accretion disk, the black hole shadow radius $R_{sh}$ increases with the observer time $t_{o}$. The ranges of the lensing ring and photon ring decrease with $t_{o}$, and the range of the direct image remains almost invariant with $t_{o}$. In the absence of an accretion disk, the shadow radius $R_{sh}$ is larger and also increases as $t_{o}$ increases.

Furthermore, we sliced the dynamical spacetime into spacelike hypersurfaces to investigate the black hole shadow for all time points ($t\geq0$). For the case with an accretion disk, the variation in $R_{sh}$ is similar to that of the apparent horizon $r_{h}$, because the inner edge of the accretion disk extends to the apparent horizon. In the absence of an accretion disk, the variation in $R_{sh}$ is similar to that of the photon sphere $r_{ps}$, because the black hole shadow boundary is determined by the photon sphere. As the variation in $r_{ps}$ is induced by the complex scalar field $\psi$, it can be stated that the variation in the size of the shadow is similarly caused by the change in $\psi$. In addition, regardless of the presence or absence of the accretion disk, the emergence of the complex scalar hair $\psi$ causes the radius $R_{sh}$ of the shadow to start changing.

In the dynamical spacetime of a black hole with a self-interacting massive complex scalar hair, we investigated the time delay $\Delta t=t_{o}-t_{e}$ of light propagating from light sources to the observer. The time delay $\Delta t$ in the direct emission from the accretion disk decreases with the increase in $L$ or $R$. The time delay $\Delta t$ in the lensing ring and photon ring first decreases and then increases as $L$ or $R$ increases. Without an accretion disk, we set a spherical background light source with $r_{s}=50$. The time delay $\Delta t$ decreases from infinity as $L$ or $R$ increases.

This study demonstrated the characteristics of dynamic black hole shadows between scenarios with and without an accretion disk. These findings not only enrich the theoretical models of dynamic black hole shadows but also provide a significant reference value in the research of observational astronomy and theoretical physics on black hole images.

\section{\bf Acknowledgments}

This work was supported by the National Natural Science Foundation of China under Grant No. 12105151, the Shandong Provincial Natural Science Foundation of China under Grant No. ZR2020QA080, and was partially supported by the National Natural Science Foundation of China under Grant No. 12375048, 11875026, 11875025 and 12035005.

\end{document}